\documentclass[fleqn,10pt]{SelfArx} 

\definecolor{color1}{RGB}{0,0,90} 
\definecolor{color2}{RGB}{0,20,20} 

\usepackage{hyperref} 
\hypersetup{hidelinks,colorlinks,breaklinks=true,urlcolor=color2,citecolor=color1,linkcolor=color1,bookmarksopen=false,pdftitle={Title},pdfauthor={Author},urlcolor=blue}

\usepackage{graphicx}
\usepackage[square]{natbib}
\usepackage{amsmath}
\usepackage{multirow}
\usepackage{subfigure}
\usepackage{pdflscape}
\usepackage{amsfonts}
\usepackage{multirow}
\usepackage{subfigure}
\usepackage{tikz}
\usetikzlibrary{calc}

\usepackage{mathtools,amssymb,lipsum}

\usepackage{cuted}
\usepackage{boondox-calo}

\usepackage{isomath}
\usetikzlibrary{calc}

\newcommand{\parderiv}[2]{\frac{\partial#1}{\partial#2}}
\newcommand{\ud}{\mathrm{d}}

\newcommand{\bs}[1]{\mathbf{#1}}

\JournalInfo{Published in IEEE Magnetics Letters, Vol. 10, 2110205, 2019} 
\Archive{\href{https://doi.org/10.1109/LMAG.2019.2956895}{DOI: 10.1109/LMAG.2019.2956895}} 

\PaperTitle{The stray- and demagnetizing field from a homogeneously magnetized tetrahedron} 

\Authors{Kaspar K. Nielsen, Andrea R. Insinga, Rasmus Bj\o{}rk} 
\affiliation{\textit{Department of Energy Conversion and Storage, Technical University of Denmark - DTU, Anker Engelunds Vej 1, DK-2800 Kgs. Lyngby, Denmark}} 
\affiliation{*\textbf{Corresponding author}: rabj@dtu.dk} 

\newcommand{\keywordname}{Keywords} 

\Abstract{The stray- and demagnetization tensor field for a homogeneously magnetized tetrahedron is found analytically. The tetrahedron is a special case of four triangular faces with constant magnetization-charge surface density, for which we also determine the tensor field. The tensor field is implemented in the open source micromagnetic and magnetostatic simulation framework MagTense and compared with the obtained magnetic field from an FEM solution, showing excellent agreement. This result is important for modeling magnetostatics in general and for micromagnetism in particular as the demagnetizing field of an arbitrary body discretized using conventional meshing techniques is significantly simplified with this approach.}

\begin{document}

\flushbottom 

\maketitle 


\thispagestyle{empty} 

\section{Introduction}
The stray- and demagnetization field tensor of a uniformly magnetized body is a fast and possibly analytical approach to calculate the magnetic field generated by a magnetized body at any point in space. The stray- and demagnetization tensor field from uniformly magnetized primitives has been derived for rectangular prisms \cite{Joseph_1965,Smith_2010}, finite cylinders \cite{Kraus_1973,Caciagli_2018}, hollow spheres \cite{Prat-Camps2016}, ellipsoids \cite{Tejedor1995}, powder samples \cite{Bleaney_1941,Bjoerk_2013,Arzbacher_2015} and an equilateral triangluar prism \cite{Tang2005}. Several such tensor fields for primitives have also been found through the Fourier space approach \cite{Beleggia2003}. Here, we present the magnetic field produced by a general triangular face with uniform magnetization and give, as a special case, the total stray- and demagnetization tensor field for a general tetrahedron, i.e. a polygon defined by four triangles that share four non-coplanar vertices. We note that the stray-field is defined as the magnetic field outside the magnetized body while the demagnetizing field is the magnetic field inside the body, both produced by the body's magnetization. The results presented in this paper are valid both inside and outside the homogeneously magnetized body.

\section{Model}
Magentostatics systems are governed by Gauss's law for magnetism and Ampere's law. Gauss's law for magnetism states that the flux density $\mathbf{B}$ is a solenoidal vector field:
\begin{equation}
\boldsymbol{\nabla} \cdot \mathbf{B} = 0\label{eq:GaussLaw}.
\end{equation}
The magnetization $\mathbf{M}$ is defined as the volume density of magnetic dipoles.  The magnetic field $\mathbf{H}$ is defined from  $\mathbf{B}$ and  $\mathbf{M}$ according to the following relation:
\begin{equation} \mathbf{B} = \mu_0(\mathbf{H} +\mathbf{M} )\label{eq:Hdef}\end{equation}
Here $\mu{}_{0}$ is the vacuum permeability.  Ampere's law is then expressed in terms of $\mathbf{H}$ and the  free current density $\mathbf{J}_{\text{\scriptsize free}}$:
\begin{equation}
\boldsymbol{\nabla} \times \mathbf{H} = \bs{J}_{\text{\scriptsize free}}.
\end{equation}


In absence of free currents, the magnetic field $\mathbf{H}$ is an irrotational vector field, i.e. its curl is zero. In these conditions it is always possible to express $\mathbf{H}$ as the gradient of a scalar field, thus guaranteeing that $\mathbf{H}$ is irrotational. We can then introduce the magnetic scalar potential $\phi_M$:
\begin{equation}
\mathbf{H}  = -\boldsymbol{\nabla}\phi_M\label{eq:PhiM}.
\end{equation}
Combining Eqs. \ref{eq:GaussLaw}, ~\ref{eq:Hdef} and ~\ref{eq:PhiM} we obtain the following equation, which is analogous to Poisson's equation:
\begin{equation}
-\boldsymbol{\nabla}{}\cdot{}(\mu{}_{0}\boldsymbol{\nabla}{}\phi_M)=-\boldsymbol{\nabla}{}\cdot{}(\mu{}_{0}\mathbf{M} )~.\label{Eq.Poisson01}
\end{equation}
 The quantity $\rho_M = -\boldsymbol{\nabla}\cdot{}\mathbf{M}$ is interpreted as the magnetic charge volume-density.
The formal solution to Eq.~\ref{Eq.Poisson01} for a homogeneously magnetized body with enclosing surface $S'$ is given by \cite{Jackson}:
\begin{equation}
    \phi_M(\bs{r})=\frac{1}{4\pi} \oint_{S'}\dfrac{\bs{\hat{n}}(\bs{r}')\cdot\bs{M}(\bs{r}')}{\|\bs{r}-\bs{r}'\|}\text{d}S'.\label{eq:PhiMsolFlat}
\end{equation}
Note that there are two sets of coordinates. The coordinates marked with a $'$ are the coordinates of the face that creates the magnetic field, whereas the non-marked coordinates are to the point at which the field is evaluated. The quantity $\sigma_M = \bs{\hat{n}} \cdot \bs{M}$ is interpreted as the magnetic charge surface-density. 
The magnetic field is obtained by combining Eq.~\ref{eq:PhiM} and Eq.~\ref{eq:PhiMsolFlat}:
\begin{equation}
\mathbf{H}(\mathbf{r})=-\boldsymbol{\nabla}\phi_M(\mathbf{r})=-\nabla\frac{1}{4\pi}\oint_{S'}\frac{\mathbf{\hat{n}}(\mathbf{r}')\cdot\mathbf{M}(\mathbf{r}')}{\|\mathbf{r}-\mathbf{r}'\|}\ud S'.
\end{equation}
Here it is worth noting that the gradient operator and the integration are with respect to different sets of coordinates: the gradient operator is with respect to the unprimed coordinates, while the integration is with respect to the primed coordinates. The order is thus interchangable.

In the following we will first derive the contribution to the magnetic field from a face shaped as a right triangle. Then we exploit that any triangle can be divided into two right triangles and apply the superposition principle to get the solution for a general triangle. We then provide a transformation necessary for obtaining the field from an arbitrarily positioned and rotated triangle from our basic solution. This, in turn, yields the contribution to the components of the stray- and demagnetization tensor field from an arbitrarily positioned and oriented triangle. Finally, we provide the details of building up a tetrahedron consisting of such four triangles and provide the total tensor field describing the stray- and demagnetizing fields from a homogeneously magnetized tetrahedron. This result has been build into the open source micromagnetic and magnetostatic simulation framework MagTense \cite{MagTense}.

\subsection{The contribution to the field from a right triangle}
\begin{figure}[!t]
\begin{tikzpicture}

	\coordinate (x1) at (4,0);
	\coordinate(x2) at (0,3);
	
	\draw[semithick,->] (-3.2,0) -- (5,0);
	\draw[semithick,->] (0,-1) -- (0,4);

	\draw[semithick,|-|] (0,-0.75) -- (4,-0.75);
	
	\draw[semithick,|-|] (0,-0.75) -- (-3,-0.75);
	
	\draw[semithick,|-|] (4.5,0) -- (4.5,3);
	
	\draw (5,-0.5) node {$x'$};
	\draw (-0.5,4) node {$y'$};
	\draw (2,-1) node {$l$};
	\draw (-2,-1) node {$k$};
	\draw (4.75,1.5) node {$h$};
	
	\draw[semithick,blue] (x1) -- (x2);
	\draw[semithick,blue] (0,0) -- (x1);
	\draw[semithick,blue] (0,0) -- (x2);
	
	\draw[semithick,violet] (-3,0) -- (x2);
	\draw[semithick,violet] (-3,0) -- (0,0);
	\draw[semithick,violet] (-0.025,0) -- (-0.025,2.95);
	
	\draw (-0.4,3) node {$\mathbf{A}'$};
	\draw (4,-0.3) node {$\mathbf{B}'$};
	\draw (-3,-0.3) node {$\mathbf{C}'$};
	\draw (-0.3,0.3) node {$\mathbf{D}'$};
	
\end{tikzpicture}
	\caption{General triangle in the local (primed) coordinate system. The triangle given by $\mathbf{A}'\mathbf{B}'\mathbf{D}'$ is the first triangle considered in the text. Please note that the lengths $h$, $k$ and $l$ are all $>0$.}
    \label{Fig_Triangle}
\end{figure}
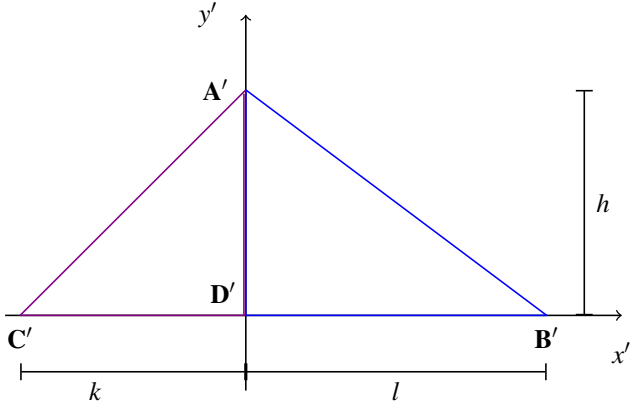

We consider a right triangle positioned in the $x'y'$-plane at $z'=0$ with the vertices $\mathbf{A}'=(l,0,0)$, $\mathbf{B}'=(0,h,0)$ and $\mathbf{D}'=(0,0,0)$ as shown in blue in Fig. \ref{Fig_Triangle}. Only the $z'$-component of the magnetization contributes to the magnetic field and the normal vector is $\mathbf{\hat{n}}'=\mathbf{\hat{z}}'$. The primes on the field and magnetization components emphasize that the field and magnetization are evaluated in the local coordinate system of the triangle as defined in Fig. \ref{Fig_Triangle}. The length of the triangle along the $x'-$axis is denoted $l$. The $x'-$component of the magnetic field is then
\begin{eqnarray}
H_x'&=&-\frac{M_z'}{4\pi}\int_0^h \ud y'\int_0^{l(1-y'/h)}\ud x'\nonumber\\
&&\parderiv{}{x}\left(\frac{1}{((x-x')^2+(y-y')^2+z^2)^{1/2}}\right)\nonumber\\
&=&N_{xz,l}'M_z'\nonumber\\
N_{xz,l}'&=&-\frac{1}{4\pi}\left(F(x,y,z,h;h,l)-F(x,y,z,0;h,l) \right.\nonumber\\
&-& \left.(G(x,y,z,h)-G(x,y,z,0))\right)
\end{eqnarray}
with
\begin{eqnarray}
F(x,y,z,y';h,l)&=&\frac{h}{\sqrt{h^2+l^2}}\mathrm{atanh}\left(\frac{f_\mathrm{n}}{f_\mathrm{d}}\right)\\
f_\mathrm{n}(x,y,z,y';h,l)&=&l^2-lx+hy-hy'\left(1+\frac{l^2}{h^2}\right)\nonumber\\
f_\mathrm{d}(x,y,z,y';h,l)&=&\sqrt{h^2+l^2}\nonumber\\
&\times&\left((x-l)^2+y^2-\frac{2(l^2-lx+hy)y'}{h}\right.\nonumber\\
&+&\left.y'^2\left(1+\frac{l^2}{h^2}\right)+z^2\right)^{1/2}\nonumber\\
G(x,y,z,y')&=&\mathrm{atanh}\left(\frac{y-y'}{\sqrt{x^2+(y-y')^2+z^2}}\right)
\end{eqnarray}

The $y-$component of the magnetic field is:
\begin{eqnarray}
H_y'&=&-\frac{M_z'}{4\pi}\int_0^l \ud x'\int_0^{h(1-x'/l)}\ud y'\nonumber\\
&&\parderiv{}{y}\left(\frac{1}{((x-x')^2+(y-y')^2+z^2)^{1/2}}\right)\nonumber\\
&=&N_{yz,l}'M_z'\nonumber\\
N_{yz,l}'&=&-\frac{1}{4\pi}\left(K(x,y,z,l;h,l)-K(x,y,z,0;h,l)\right.\nonumber\\
&-&\left.(L(x,y,z,l)-L(x,y,z,0)\right)
\end{eqnarray}
with
\begin{eqnarray}
K(x,y,z,x';h,l)&=&\frac{l}{\sqrt{h^2+l^2}}\mathrm{atanh}\left(\frac{k_\mathrm{n}}{k_\mathrm{d}}\right)\nonumber\\
k_\mathrm{n}(x,y,z,x';h,l)&=&h^2-hy+lx-lx'\left(1+\frac{h^2}{l^2}\right)\nonumber\\
k_\mathrm{d}(x,y,z,x';h,l)&=&\sqrt{h^2+l^2}\nonumber\\
&\times&\left((y-h)^2+x^2-\frac{2x'(h^2+lx-hy)}{l}\right.\nonumber\\
&+&\left.x'^2\left(1+\frac{h^2}{l^2}\right)+z^2\right)^{1/2}\nonumber\\
L(x,y,z,x')&=&\mathrm{atanh}\left(\frac{x-x'}{\sqrt{(x-x')^2+y^2+z^2}}\right)
\end{eqnarray}

The $z-$component of the field becomes:
\begin{eqnarray}
H_z'&=&-\frac{M_z'}{4\pi}\int_0^l \ud x'\int_0^{h(1-x'/l)}\ud y'\nonumber\\
&&\parderiv{}{z}\left(\frac{1}{((x-x')^2+(y-y')^2+z^2)^{1/2}}\right)\nonumber\\
&=&N_{zz,l}'M_z'\nonumber\\
N_{zz,l}'&=&-\frac{1}{4\pi}\left(P(x,y,z,l;h,l)-P(x,y,z,0;h,l)\right.\nonumber\\
&-&\left.(Q(x,y,z,l)-Q(x,y,z,0))\right)
\end{eqnarray}
with
\begin{eqnarray}
P(x,y,z,x';h,l)&=&\mathrm{atan}\left(\frac{p_\mathrm{n}}{p_\mathrm{d}}\right)\nonumber\\
p_\mathrm{n}(x,y,z,x';h,l)&=&x(h-y)-x'(h(1-\frac{x}{l})-y)-\frac{h(x^2+z^2)}{l}\nonumber\\
p_\mathrm{d}(x,y,z,x';h,l)&=&z\left((y-h)^2+x^2+x'^2\left(1+\frac{h^2}{l^2}\right)\right.\nonumber\\
&-&\left.\frac{2x'(h^2+lx-hy)}{l}+z^2\right)^{1/2}\nonumber\\
Q(x,y,z,x')&=&-\mathrm{atan}\left(\frac{(x-x')y}{z\sqrt{(x-x')^2+y^2+z^2}}\right).
\end{eqnarray}
Some of these integrals were evaluated using the Rule-based integrator \cite{Rich2018}. In order to get the contribution from an entire triangle, which can always be decomposed into two right triangles, as shown in Fig. \ref{Fig_Triangle}, the tensor components of these right triangles are added such that $N_{xz,\mathrm{tot}}'=N_{xz,k}'+N_{xz,l}'$, $N_{yz,\mathrm{tot}}'=N_{yz,k}'+N_{yz,l}'$ and $N_{zz,\mathrm{tot}'}=N_{zz,k}'+N_{zz,l}'$. The length along $x'$ for the right triangle $\mathbf{A}'\mathbf{C}'\mathbf{D}'$ in Fig. \ref{Fig_Triangle} is denoted $k$,  where it is noted that the integration limits on the integrals over $\ud x'$ are switched. The lower limit is thus $-k$ when the integral over $\ud x'$ is the outer integral and $-k(1-y'/h)$ when this integral is the inner. It is further noted that $\mathbf{C}'=(-k,0,0)$. As an example, for the triangle in the second quadrant in Fig. \ref{Fig_Triangle} we have:
\begin{eqnarray}
N_{xz,k}'&=&-\frac{1}{4\pi}(G(x,y,z,h)-G(x,y,z,0) \nonumber\\
&-& (F(x,y,z,h;h,k)-F(x,y,z,0;h,k))).
\end{eqnarray}

\subsection{Numerical Singularities}
The closed expressions given above for the local coordinate system demagnetization tensor field components contain some numerical singularities. It is clear that it is a requirement that $h,k,l>0$ otherwise the triangular face assumption breaks down. Furthermore, the solutions for $H_x'$, $H_y'$ and $H_z'$ are numerically singular along the $x-$ and $y-$ axes for $y=z=0$ and $z=0$, respectively, in the local system as well as right on the lines $A'B'=A'-B'$ and $B'C'=B'C'$. However, both field components can be obtained on these axes and lines by chosing a suitably small $z$ value for the result to be numerically stable. At $z=0$ the local demagnetization tensor component $N_{zz,l}'$ is zero as the functions $P(x,y,z,x')$ and $Q(x,y,z,x')$ given above cancel out even though the arguments to the atan function are infinite since the function arguments approach infinity at the same rate.

\subsection{Arbitrarily positioned and oriented triangular face}
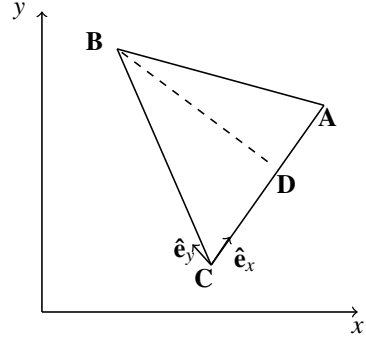
\begin{figure}[!t]
\begin{tikzpicture}

	\coordinate (x1) at (1.25,1.75);
	\coordinate(x2) at (-1.5,2.5);
	
\draw[semithick,->] (-2.5,-1) -- (1.7,-1);
\draw[semithick,->] (-2.5,-1) -- (-2.5,3);

	
	
\draw (1.7,-1.2) node {$x$};
\draw (-2.8,3) node {$y$};
	
	\draw[semithick] (x1) -- (x2);
	\draw[semithick,dashed] (0.5,1) -- (x2);
	
	\draw[semithick] (-0.25,-0.375) -- (x2);
	\draw[semithick] (-0.25,-0.375) -- (x1);
	\draw[semithick,->] (-0.25,-0.375) -- (0,0); 
	\draw[semithick,->] (-0.25,-0.375) -- (-0.5,-0.1); 

	\draw (1.3,1.6) node {$\mathbf{A}$};	
	\draw (-1.8,2.6) node {$\mathbf{B}$};
	\draw (-0.35,-0.55) node {$\mathbf{C}$};
	\draw (0.75,0.7) node {$\mathbf{D}$};
	
	\draw (0.2,-0.3) node {$\mathbf{\hat{e}}_x$};
		\draw (-0.6,-0.2) node {$\mathbf{\hat{e}}_y$};
	
\end{tikzpicture}
	\caption{General triangle in the global coordinate system. Note that the unit vector in the $z-$direction points out of the plane.}
    \label{Fig_GeneralTriangle}
\end{figure}

We now consider a general triangular face in Euclidian geometry characterized by three points, $\mathbf{A},\ \mathbf{B}$ and $\mathbf{C}$ in the global coordinate system, G. It is a requirement that the three vertices are not collinear.

The goal is to find the linear transformation from this general triangle to a local coordinate system, L, with the triangle placed as shown in Fig. \ref{Fig_Triangle}. This transformation is represented by the matrix $P^{-1}$. Its inverse, $P$, is the matrix representing the transformation from the local to the global coordinate system and this consists of the three column unit vectors that form the orthonormal basis of the local coordinate system in terms of global coordinates:
\begin{eqnarray}
P=\left\{[\mathbf{\hat{e}}_x]_G^\top,[\mathbf{\hat{e}}_y]_G^\top,[\mathbf{\hat{e}}_z]_G^\top\right\}.
\end{eqnarray}
As $P$ is orthogonal then $P^{-1}=P^\top$. Before applying this transformation the triangle is shifted such that the point $\mathbf{D}$ on the triangle (see Fig. \ref{Fig_GeneralTriangle}) coincides with $\mathbf{D}'$ in the local coordinate system, i.e.
\begin{eqnarray}
\begin{bmatrix}
(\mathbf{A}')^\top  & (\mathbf{B}')^\top	& (\mathbf{C}')^\top
\end{bmatrix}&&\nonumber\\
=
P^{-1}
\begin{bmatrix}
(\mathbf{A}-\mathbf{D})^\top & (\mathbf{B}-\mathbf{D})^\top & (\mathbf{C}-\mathbf{D})^\top
\end{bmatrix}.&&
\end{eqnarray}
Without a loss of generality, we assume that $\mathbf{B}$ is the point with the largest angle in the general triangle, i.e. $\angle \mathbf{B}=\mathrm{max}\{\angle \mathbf{A},\angle \mathbf{B},\angle \mathbf{C}\}$. In this way we ensure in the local coordinate system that the triangle may always be split in two right triangles as per Fig. \ref{Fig_Triangle}. Assuming that the normal vector to the triangular face is $[\mathbf{\hat{e}}_z]_G=\frac{\mathbf{AC}\times\mathbf{BC}}{\|\mathbf{AC}\|\|\mathbf{BC}\|}$ and defining the first unit vector as $[\mathbf{\hat{e}}_x]_G=\frac{\mathbf{AC}}{\|\mathbf{AC}\|}$ the second unit vector becomes $[\mathbf{\hat{e}}_y]_G=[\mathbf{\hat{e}}_z]_G\times[\mathbf{\hat{e}}_x]_G$; see Fig. \ref{Fig_GeneralTriangle}. All that remains now is to find the point $\mathbf{D}$, which may be done via trigonometry: $\mathbf{D}=\|\mathbf{BC}\|\cos\left(\angle \mathbf{C}\right)[\mathbf{\hat{e}}_x]_G+\mathbf{C}$.

With the point $\mathbf{D}$ determined, the translation and transformation can be fully calculated, and the magnetic field from a triangular face as defined above and assuming a magnetization vector $\mathbf{M}=(M_x,M_y,M_z)$ is thus given by:
\begin{eqnarray}
\mathbf{H}(\mathbf{r})=P\tensorsym{N}'(P^{-1}(\mathbf{r}-\mathbf{D}))P^{-1}\mathbf{M}^\top.\label{eq_H_tri_face}
\end{eqnarray}
The partial demagnetization tensor field in the local coordinate system is given by:
\begin{eqnarray}
\tensorsym{N}'(\mathbf{r})=
\begin{bmatrix}
0			&	0	&	N_{xz,l}'+N_{xz,k}'\\
0			&	0	&	N_{yz,l}'+N_{yz,k}'\\
0			&	0	&	N_{zz,l}'+N_{zz,k}'
\end{bmatrix}.
\end{eqnarray}
\subsection{The field from a homogeneously magnetized tetrahedron}
Given four vertices $\mathbf{v}_i$, $i=1..4$ that are not co-planar (and thus also not collinear), we can define the four triangles that make up a tetrahedron with these four vertices. The total field is then the sum of the contributions of each triangular face as may be found above. The only remaining point is to ensure the normal vector of each triangle to point outwards and this is done as follows. Assume the vertices $\mathbf{v}_i$ where $i=1,2,3$ form a triangular face and that they are ordered such that $\mathbf{v}_2$ is the vertex at the largest angle in the triangle. It is also assumed that $\mathbf{v}_1$ corresponds to $\mathbf{A}$ as above, $\mathbf{v}_2$ to $\mathbf{B}$ and $\mathbf{v}_3$ to $\mathbf{C}$. Then if
\begin{eqnarray}
((\mathbf{v}_1-\mathbf{v}_3)\times (\mathbf{v}_2-\mathbf{v}_3)) \cdot (\mathbf{v}_4-\mathbf{v}_3)>0,\nonumber
\end{eqnarray}
where $\mathbf{v}_4$ is the tetrahedron vertex that is not part of the triangular face considered, the sign of the normal should be changed, which amounts to interchanging $\mathbf{v}_1$ and $\mathbf{v}_3$.

The change-of-basis transformation matrix, $P$, may be written in terms of $\mathbf{v}_{1..3}$, assuming the vertices to be ordered, as:
\begin{eqnarray}
P_{123}&=&\frac{1}{\|(\mathbf{v}_1-\mathbf{v}_3)\|}\left.\Bigg\{\right.\nonumber\\
&&\left.(\mathbf{v}_1-\mathbf{v}_3)^\top; \left(\frac{((\mathbf{v}_1-\mathbf{v}_3)\times(\mathbf{v}_2-\mathbf{v}_3))\times(\mathbf{v}_1-\mathbf{v}_3)}{\|(\mathbf{v}_1-\mathbf{v}_3)\|\|(\mathbf{v}_2-\mathbf{v}_3)\|}\right)^\top;\right.\nonumber\\
&&\left. \left(\frac{(\mathbf{v}_1-\mathbf{v}_3)\times(\mathbf{v}_2-\mathbf{v}_3)}{\|(\mathbf{v}_2-\mathbf{v}_3)\|}\right)^\top\right\}.
\end{eqnarray}
The total field at the point $\mathbf{r}$ is then found through summation over the contributions from four triangles given by Eq. \ref{eq_H_tri_face} with
\begin{eqnarray}
\mathbf{D}_{123}&=&\mathbf{v}_3 + (\mathbf{v}_1-\mathbf{v}_3)\nonumber\\
&\times&\frac{\|(\mathbf{v}_1-\mathbf{v}_3)\|^2+\|\mathbf{v}_2-\mathbf{v}_3\|^2-\|\mathbf{v}_2-\mathbf{v}_1\|^2}{2\|(\mathbf{v}_1-\mathbf{v}_3)\|^2}.\nonumber\\
\end{eqnarray}
The contributions from the remaining three triangles are found through cyclic permutation of the indices (including, of course, $\mathbf{v}_4$ in order to obtain $P_{412}$ and $\mathbf{D}_{412}$ etc.).

\section{Verification}
The above calculations of the stray- and demagnetization tensor field can be verified by comparison with a finite element method computation of the magnetic flux density from a tetrahedron. The magnetic field is here computed using the finite element software Comsol, which in a finite element framework solves Eq.~\ref{Eq.Poisson01} with respect to $\phi_M$ for given magnetization $\mathbf{M}$.

\begin{figure}[!t]
  \centering
  \includegraphics[width=1\columnwidth]{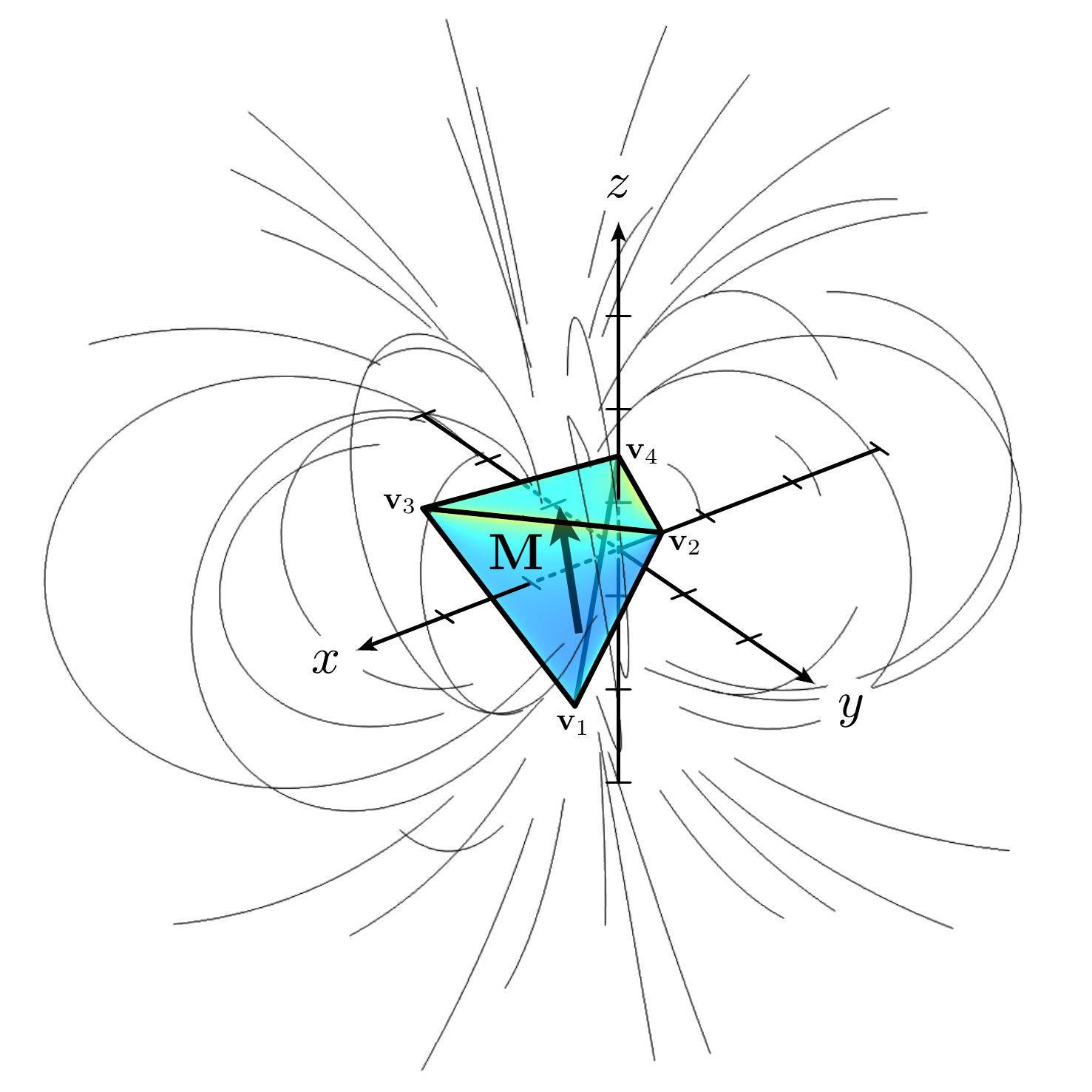}
  \caption{The tetrahedron used for the numerical verification. The magnetization vector $\mathbf{M}$ and the field lines of $\mathbf{H}$ are shown. The field profiles of Fig.~\ref{Fig_Tetrahedron_Comsol} correspond to the Cartesian axes shown here, i.e. through the point $(x,y,z) = (3, 3, 2.5)$. The ticks shown on the axes are spaced by $1$~mm intervals.}
  \label{Fig_Tetrahedron_3D}
\end{figure}
\begin{figure}[!t]
  \centering
  \includegraphics[width=1\columnwidth]{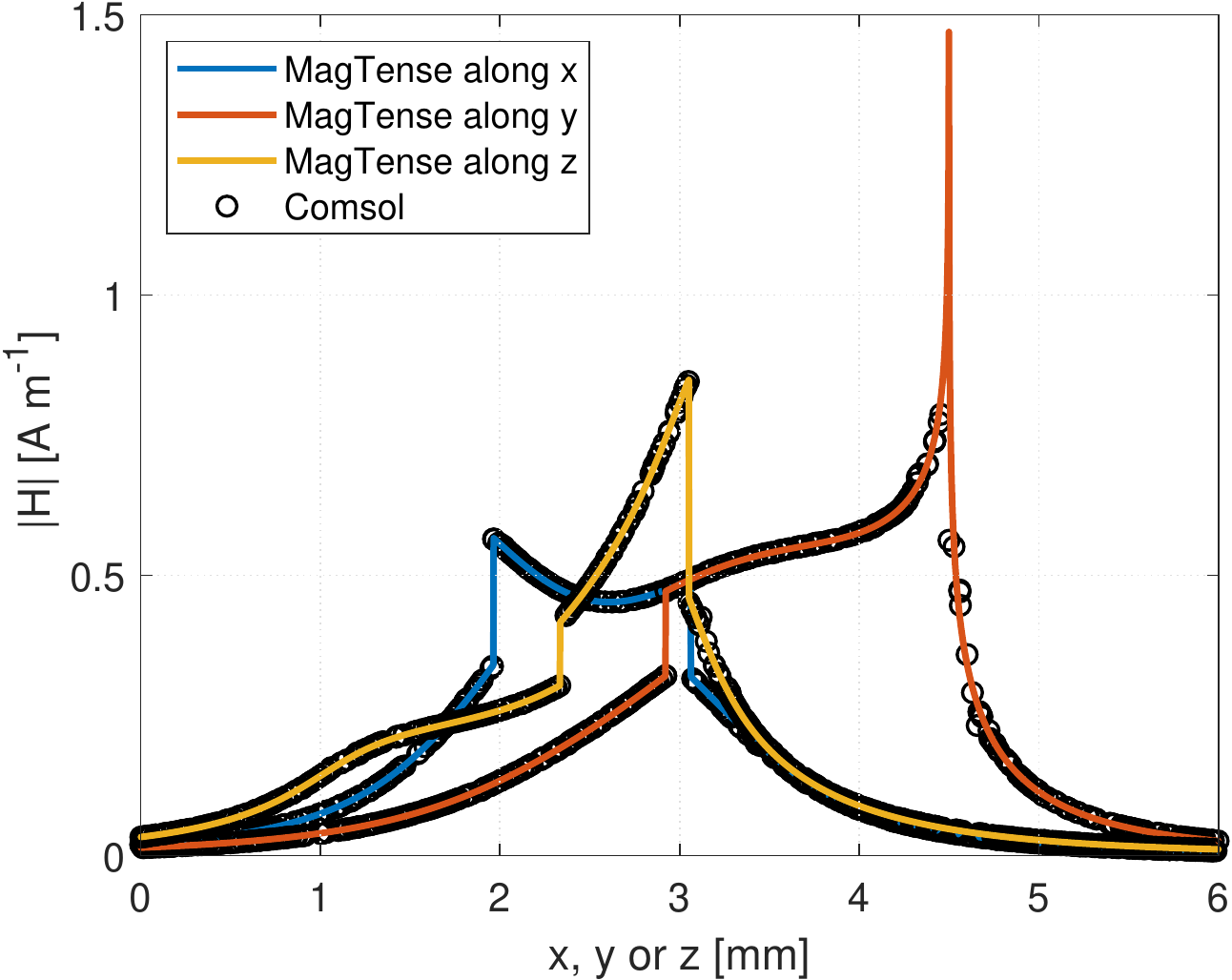}
  \caption{The norm of the magnetic field as calculated using the components of the tensor field given in the text using MagTense and as calculated using the FEM framework Comsol, for the example geometry shown in Fig.~\ref{Fig_Tetrahedron_3D} along each of the Cartesian axes through the point $(x,y,z) = (3, 3, 2.5)$.}
  \label{Fig_Tetrahedron_Comsol}
\end{figure}

We consider the tetrahedron shown in Fig.~\ref{Fig_Tetrahedron_3D} with the following four vertices: $(x,y,z)=\mathbf{v}_1=(2.5, 3, 1)$, $\mathbf{v}_2=(2, 1, 4)$, $\mathbf{v}_3=(1.5, 4, 3)$, and $\mathbf{v}_4=(4.5, 5, 2)$~mm.
The tetrahedron has a magnetization of $\mathbf{M}=(0.32, 0.74, 0.89)$~A$/$m. The norm of the magnetic field is calculated along each of the Cartesian axes through the point $(x,y,z) = (3, 3, 2.5)$~mm inside the tetrahedron and is shown in Fig. \ref{Fig_Tetrahedron_Comsol}. As can be seen form the figure there is an excellent agreement between Comsol and the demagnetization tensor approach discussed here. This example is part of the verification examples for MagTense and is available online \cite{MagTense}.

\section{Conclusion}
The stray- and demagnetization tensor field was found for a right triangle in the $x'y'-$plane followed by the tensor field for a general triangle arbitrarily positioned and oriented in space. With the proper change of basis, the total stray- and demagnetization tensor field for a homogeneously magnetized tetrahedron was obtained from the right triangle solution. The code for numerically evaluating this tensor field for an arbitrary tetrahedron is published in the micromagnetism and magnetostatics framework MagTense and is available online as Open Source in both a Matlab and a Fortran version \cite{MagTense}. As the computations involved are very fast ($<3\ \mu\mathrm{s}$ on a normal PC per tetrahedron in the Fortran implementation) the solution provided here has a great potential if used in combination with conventional meshing techniques for enabling fast and consistent computation of the stray- and demagnetizing field of a magnetized body with locally varying magnetization. In such a computation each tetrahedron in the mesh would be assumed to be homogeneously magnetized but not with the same magnetization. The geometrical part of the problem, i.e. finding the tensor field as described above, needs only to be done once for a given problem / geometry. Subsequently, finding the self-consistent solution, i.e. the set of magnetization vectors that satisfy the constitutive relation between magnetic field and magnetization assumed, may be done very fast.

The special case of the tetrahedron may of course be extended directly to the case of a general polygon that may always be split into triangles and to any homogeneously magnetized body who's enclosing surface is approximated by such a polygon.

\section*{Acknowledgements}
This work was financed partly by the Energy Technology Development and Demonstration Program (EUDP) under the Danish Energy Agency, project no. 64016-0058, partly by the Danish Research Council for Independent Research, Technology and Production Sciences projects no. 7017-00034 and 8022-00038 and partly by the Poul Due Jensen Foundation project on Browns paradox in permanent magnets, project no. 2018-016.

\end{document}